# Ronchigram Simulation and Training through Ronchigram.com

Suk Hyun Sung[1], Noah Schnitzer[2], William Millsaps[3], Lena F. Kourkoutis[4,5], Robert Hovden[1]

[1]Department of Materials Science and Engineering, University of Michigan, Ann Arbor, MI 48109
[2]Department of Materials Science and Engineering, Cornell University, Ithaca, NY 14853
[3]Engineering Physics Program, University of Michigan, Ann Arbor, MI 48109
[4]School of Applied and Engineering Physics, Cornell University, Ithaca, NY 14853
[5]Kavli Institute at Cornell for Nanoscale Science, Cornell University, Ithaca, NY 14853

*nis29@cornell.edu

## Abstract

This article introduces a training simulator for electron beam alignment using Ronchigrams. The interactive web application, www.ronchigram.com, is an advanced educational tool aimed at making scanning transmission electron microscopy (STEM) more accessible and open. For experienced microscopists, the tool offers on-hand quantification of simulated Ronchigrams and their resolution limits.

## Introduction

Ronchigrams are at the heart of atomic-resolution imaging with a scanning transmission electron microscope (STEM) [1,2]. The development of aberration correction [3–5] propelled widespread use of STEM for materials, chemistry, and physics [6–10], however, the sophisticated electromagnetic lenses have made electron beam alignment more complex, and better understanding of Ronchigrams is now required. The quality and interpretability of STEM data are determined by the ability to analyze nuanced aberrations hidden within the baroque structure of a Ronchigram [11,12]. Unfortunately, assessing a Ronchigram (Figure 1) has a steep learning curve, and gaining the necessary experience requires costly microscope time.

Ronchigram.com is a new simulator for rapid training of microscopists and provides advanced analytics for the experienced. In this tutorial, we will introduce the Ronchigram.com training interface and then use it to explore different types of aberrations. Training exercises will build familiarity with the interface and provide a feel for experimental aberration correction. As the Ronchigram dynamically changes with the input, you can simultaneously inspect the electron probe in real space.

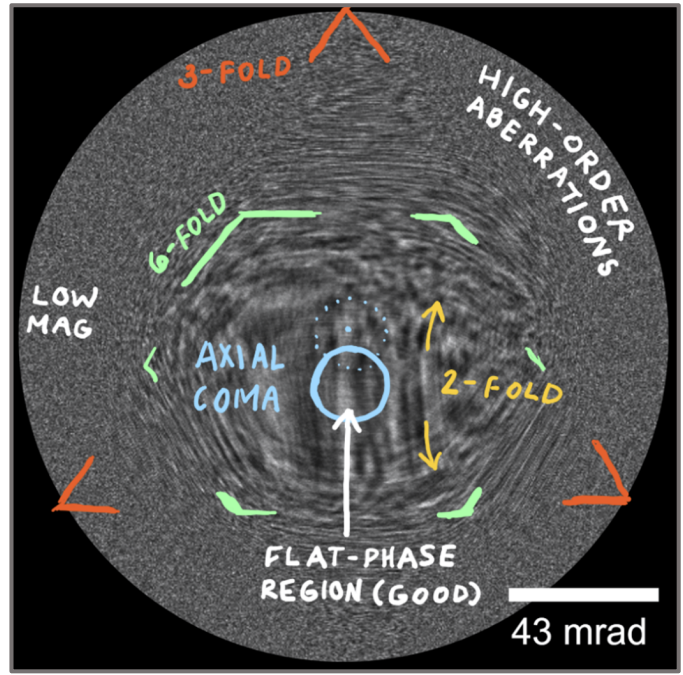

**Figure 1: Ronchigrams have symmetries and features that reveal the lens aberrations present.** Above, a simulated Ronchigram at 300 keV with a 95 mrad aperture semi-angle shows 2-fold (yellow), 3-fold (red), and 6-fold (green) symmetries along with a center shift due to axial coma (blue). With training, the corresponding aberrations can be identified and corrected.

**The Updated Ronchigram.com (Version 2.0)**

Ronchigram.com runs locally in a web browser—no installation is required—and runs on any platform (macOS, Windows, Linux) including cell phones (for example, iOS, Android). An offline mode can also be installed to execute Ronchigram.com outside of a web browser, which is particularly valuable for running simulations on microscope computers without an internet connection. Since our previous introduction to Ronchigrams [13], additional functionality has been added to the tool.

The latest version of Ronchigram.com (v2.0) contains several new features: a training module helps build intuition for beam alignment (for example, stigmation, focus, coma); visualization of the corresponding real-space probe (the point spread function) is simultaneously displayed; convergence angle is now calculated from the Strehl ratio, which optimizes probe size better than traditional π/4 phase rules [14]; and other features such as documentation, higher sampling, lower latency, and improved graphic design are included.

**1.1 Tutorial: Assessing Aberrations with the Ronchigram**

To begin alignment training, open Ronchigram.com in the *training mode* by visiting ronchigram.com/vasco (Figure 2).

For optimal resolution, the electron beam intensity in real space (the probe function) has to be as tightly focused (that is, sharply peaked with minimal tails) and symmetric (that is, round) as possible. Achieving an optimal probe requires that aberrations be minimized across the largest aperture allowed [15]. Experimentally, the Ronchigram reveals the structure and magnitude of aberrations in the objective lens [16]. Here we will observe how aberrations affect the Ronchigram and the electron beam shape.

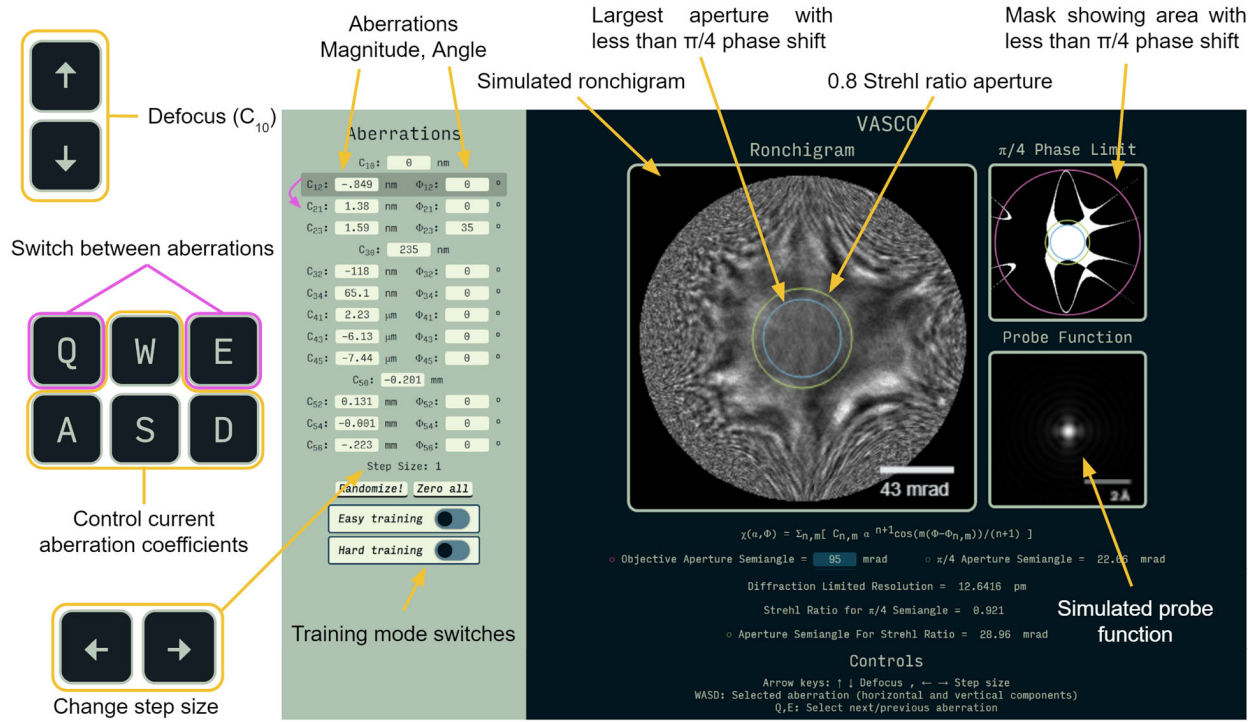

**Figure 2: Ronchigram.com in training mode.** The left panel lists up to 5th-order geometric aberrations. The right panel displays the Ronchigram alongside a map of areas where the aberration function magnitude is below π/4. The green and blue circles show optimal aperture placement. The real space electron probe is displayed lower right. Keyboard inputs (WASD) adjust the aberration term highlighted by the shaded box. Q and E change the selected term. Up and down arrow keys are mapped to defocus; left and right arrow keys change the step size. The training modes (Easy and Hard) randomize the first three terms ($C_{10}$, $C_{12}$, $C_{21}$) and hide all coefficients, which the user corrects. Hard training mode also hides the phase shift and probe function—mimicking experimental settings. Users are notified when alignment is achieved. The "Randomize!" button generates new aberration values. The "Zero all" button sets all coefficients to zero.

### 1.2 Ronchigram and Aberration Basics

The objective aperture size is a critical parameter for enhancing the diffraction-limited resolution, but a larger aperture comes with a trade-off: better resolution also requires a larger aberration-free ("flat-phase") region [11]. Figure 3e shows a green circle marking the flat-phase region, and, consequently, the largest allowed aperture after alignment. The objective of any microscopist is to tune the Ronchigram to produce the largest flat-phase region possible and place an aperture centered within it to set the final resolution. Regions of the Ronchigram with rapid variation should not be included within the aperture, as these aberrated portions degrade probe shape and resolution.

Ronchigram.com uses two methods to automatically calculate the maximum aperture allowed in a Ronchigram: the blue circle marks the aperture semi-angle suggested by the π/4 heuristic, and the green circle marks the semi-angle suggested by the 0.8 Strehl ratio heuristic. The 0.8 Strehl ratio heuristic better optimizes probe size when non-spherical aberrations are present [14]. The π/4 heuristic is popular, but less accurate at predicting optimal aperture size.

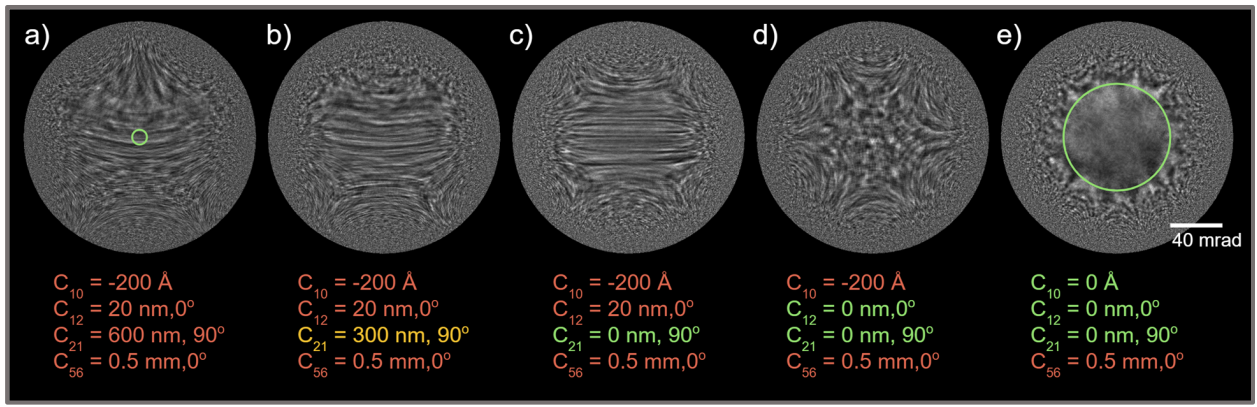

**Figure 3: The features of a Ronchigram can be recognized and tuned.** Above, five simulated Ronchigrams at 300 keV with a 95 mrad aperture semi-angle show what this process looks like. a) The Ronchigram begins with defocus, 2-fold astigmatism, axial coma, and 6-fold symmetry. b) Axial coma is reduced. c) Axial coma is fully corrected, and 2-fold astigmatism dominates. d) 2-fold astigmatism is corrected, and the microscope can be adjusted to proper focus. e) A well-aligned Ronchigram has a large low-aberration region (green circle), though higher-order aberrations remain farther out from the center.

Figure 3 shows several distinct features of lower-order aberrations that electron microscopists are expected to recognize and correct. In Figure 3a, the most featureless "flat" region is not in the middle of the Ronchigram, which is a sign of some form of axial coma. In Figure 3c, the straight lines are a strong sign of 2-fold astigmatism and defocus. In Figure 3d, the Ronchigram is not at proper focus and has higher-order aberrations around the edge. When focused, as in Figure 3e, the corrected Ronchigram has a large, flat, featureless region inside of the colored circle showing the microscopist the optimal aperture semi-angle. Note that higher-order aberrations will always remain farther out from the center.

The following exercises will help to build an understanding of what the different lower-order aberrations look like in a Ronchigram.

1. **Defocus ($C_{10}$)**. Starting with the default Ronchigram, change the defocus, C10, through the range of -20 nm to +20 nm using the arrow keys. Pay attention to the flat region in the center and the optimal aperture (colored circle) as they change size.

2. **2-Fold Astigmatism ($C_{12}$)**. Leave the defocus between 15 and 20 nm. Use the keys (WASD) to change the magnitude of $C_{12}$ to about 20 nm, and observe the characteristic streaks, or lines, that appear in the middle of the Ronchigram. Also, observe how the directionality of the streaks in the Ronchigram changes when under- and over-focused (for example, -20 nm and +20 nm $C_{10}$). 2-fold astigmatism is an important aberration to recognize and is generally corrected manually.

3. **Axial Coma ($C_{21}$)**. Set $C_{12}$ to 0 nm and $C_{10}$ to 20 nm, then increase axial coma to ~200nm. The Ronchigram bends, and the central flat region is pushed to the side. Now shift the focus from 20 nm to -20 nm and observe as the Ronchigram changes. The key to identifying axial coma is that the flat region of the Ronchigram seems to swing across the aperture; the direction of the swing corresponds to the direction of the coma. To correct $C_{21}$, minimize this swing while going through focus.

4. **Higher-Order Terms**. Briefly look at some higher-order terms. 3-fold astigmatism ($C_{23}$), spherical aberration ($C_{30}$), and star aberration ($C_{32}$) all have distinctive features but are hard to distinguish among other higher-order aberrations. To view them without interference, press the "zero all" button, and set the coefficient of choice to an arbitrary value.

**1.3 Aberration Correction Training**

The following exercise will help build familiarity with the training mode and how it works.

1. **Setup:** To begin, press the "Zero all" button, and then switch "Easy training" to on. By default, this will randomize the first three terms ($C_{10}$, $C_{12}$, & $C_{21}$) and then hide them from the user.

2. **Tuning:** Adjusting $C_{10}$, $C_{12}$, & $C_{21}$, use your understanding of basic aberrations to make the Ronchigram as flat as possible, as in Figure 3, where the "flat" or "smooth" region shown inside of the green circle becomes larger. The goal is to get the $\pi/4$ aperture semi-angle (blue circle) to at least 20 mrad.

3. **Feedback:** If you get stuck, you can peek at the offset coefficients by switching the training mode to off. The goal is to get the coefficients close to zero. Note that when training mode is re-enabled, the coefficients will be randomized again.

4. **Introducing higher order aberrations :** In steps 1, 2, and 3, only three coefficients were randomized, making your job easier, but also less realistic. When tuning a real microscope, there will be higher-order aberrations. Though you primarily correct only the first 3 terms on a real STEM, Ronchigrams will look significantly different than in previous steps.

   To provide a realistic experimental simulation, simply press "Randomize!" while in training mode. This will produce a Ronchigram with significant $C_{10}$, $C_{12}$, and $C_{21}$ aberrations alongside realistic higher-order aberrations.

5. **Hard mode - optional:** The "Hard training" switch removes the phase map and probe function, leaving only the Ronchigram, just like in experimental conditions. To prepare for tuning a real microscope, use this mode with randomized higher-order aberrations. At the bottom of the page, there is also a box that allows you to select "Correction training terms." To increase difficulty further (beyond what you'll normally have to do to tune), change this number to a higher value (for example, 5), and enter training mode.

**Conclusion**

Ronchigram.com is an educational tool aimed at making advanced electron microscopy more accessible and open. Simulator and training tools not only lower the cost but improve the level of operation to maximize the quality and success of scientific experiments. Experienced users will find it to be a quick and valuable reference.

**Acknowledgements**

Ronchigram.com is an educational, open-source project released under the 3-clause BSD license, free for non-commercial use. The release is available on GitHub at https://github.com/sukhsung/vasco-app/. This tutorial was originally developed for the 2021 Platform for the Accelerated Realization, Analysis, and Discovery of Interface Materials (PARADIM) summer school on electron microscopy at Cornell University supported by the National Science Foundation under Cooperative Agreement No. DMR-2039380. R.H. recognizes support from ARO grant no. W911NF-22-1-0056. N.S. was supported by the NSF Graduate Research Fellowship (DGE-2139899). Authors thank Abha Dabak-Wakankar for feedback.